\documentclass [] {aa} 
\usepackage{epsfig, graphicx,longtable}
\usepackage{natbib}
\usepackage{times}

\bibpunct{(}{)}{;}{a}{}{,}

\begin{document}


\title{The CORALIE survey for southern extra-solar
  planets\subtitle{XI. The return of the
  giant planet orbiting HD\,192263\thanks{Based on observations collected at the La Silla
    Observatory, ESO (Chile), with the {\small CORALIE} spectrograph
    at the 1.2-m Euler Swiss telescope and at the La Palma
    Observatory, Spain, with the P7 photometer at the 1.2-m {\small
      MERCATOR} Belgian telescope}}} 

\author{N.C.~Santos\inst{1,2} \and 
        S.~Udry\inst{1} \and 
        M.~Mayor\inst{1} \and 
	D.~Naef\inst{1} \and 
	F.~Pepe\inst{1} \and 
	D.~Queloz\inst{1} \and 
	G.~Burki\inst{1} \and 
	N.~Cramer\inst{1} \and
	B.~Nicolet\inst{1} 
	} 

\offprints{Nuno C. Santos, \email{Nuno.Santos@oal.ul.pt}}

\institute{
	Observatoire de Gen\`eve, 51 ch.  des 
	Maillettes, CH--1290 Sauverny, Switzerland 
	\and
	Centro de Astronomia e Astrof{\'\i}sica da Universidade de Lisboa,
        Observat\'orio Astron\'omico de Lisboa, Tapada da Ajuda, 1349-018 
        Lisboa, Portugal
        } 

\titlerunning{} 


\abstract{ The presence of a planet around the K dwarf
  \object{HD\,192263} was recently called into question by the
  detection of a periodic photometric signal with the same period as
  the one observed in radial velocity.  In this paper, we investigate
  this possibility, using a combination of radial-velocity,
  photometry, and bisector measurements obtained simultaneously. The
  results show that while the observed radial-velocity variation is
  always very stable in phase, period, and amplitude, the photometric
  signal changes with time.  The combined information strongly
  suggests that the observed radial-velocity variation is being
  produced by the presence of a planet, as firstly proposed. The
  photometric variations are either not connected to the planetary
  companion, or can eventually be induced by the interaction between
  the planet and the star. Finally, the radial-velocity data further
  show the presence of a long term trend, whose origin, still not clear, 
  might be related to the presence of another companion to the system. 
  \keywords{planetary systems -- stars: individual: HD\,192263} }

\maketitle

\section{Introduction}

Radial-velocity techniques have brought to light more than 100
planetary candidates around solar-type stars\footnote{See e.g.
  http://obswww.unige.ch/Exoplanets}.  The odd properties of many of
the exoplanet candidates raised some scepticism from the community.
The first exoplanet discovered, orbiting the solar-type star 51\,Peg
\citep{May95}, is itself a good example. Its particularly short-period
orbit ($\sim$4.23 days) led some astronomers to cast doubts about its
existence: e.g. \citet{Gra97} suggested that the radial-velocity
variations were due to non-radial pulsations rather than to the
presence of a planetary mass companion. Later on, this result was
withdrawn \citep{Gra98} and the presence of the planet around 51\,Peg
confirmed.

Other similar examples exist in the literature. Exploring the fact
that the radial-velocity technique only gives us the minimum mass for
the companion, \citet{Han01} suggested that the planetary candidates
were in fact low mass stars on orbits seen edge-on.  This result was
easily refuted by statistical arguments in the case of random
orbital-plane inclinations \citep{Hal00,Jor01,Pou01,PouA01}.  Again,
the ``planetary origin'' of the radial-velocity variations was then
considered to be the best one.

It is known that the radial-velocity technique is not sensitive only
to the motion of a star around the center of mass of a star/planet
system. Intrinsic variations, such as non-radial pulsation
\citep{Bro98}, inhomogeneous convection or spots, are expected to
induce radial-velocity variations
\citep[e.g.][]{Saa97,Saa98,San00b,Pau02,Tin02}.  These situations can
prevent us from finding planets (if the perturbation is larger than
the orbital radial-velocity variation) or give us false candidates (if
they produce a periodic signal over a few rotational periods).  A good
example of this effect is given by the periodic radial-velocity
variation observed for the dwarf {\object HD\,166435}, that was shown
to be due to a spot rather than to the presence of a planetary
companion \citep{Que01}.

The presence of unknown stellar blends can also induce spurious
radial-velocity signals, which can ``simulate'' the presence of a
planetary companion in the case of triple systems. An example is given
by HD\,41004 in which the moving spectrum of a faint spectroscopic
binary companion induces a planetary-type signature on the primary
star \citep{San02,Zuc03}.

In this context, another planetary companion that was recently called
back into question is the case of \object{HD\,192263}. The star was
announced to harbor a Jupiter-mass planetary companion on a
$\sim$24-day period orbit \citep{San00a,Vog00}. Recently however,
\citet{Hen02} have detected a photometric variation with a period
compatible with the period observed in the radial-velocity data.  The
authors have then concluded that the planet around this star was no
longer needed to explain the radial-velocity signal, and that the case
of \object{HD\,192263} was similar to the one observed in
\object{HD\,166435}.

In this study, we analyze in more details the situation concerning
\object{HD\,192263}. Long term and simultaneous radial-velocity,
bisector, and photometric measurements are presented\footnote{The radial-velocity and photometry measurements will be available 
in electronic form at the CDS}. The results show
that the presence of a planet is still needed to explain the observed
radial-velocity signal. The sporadic observed photometric variations
can eventually be explained as the result of interactions between the
planet and the host star. In Sect.\,\ref{parameters} we review the
stellar parameters of \object{HD\,192263}, and in Sect.\,\ref{rv} we
present the available radial-velocity data. In Sect.\,\ref{spurious}
we analyze the combined radial-velocities, bisector, and photometric
data, exploring the different possibilities to explain the
observations. We conclude in Sect.\,\ref{conclusions}.

\section{Stellar Parameters}
\label{parameters}

The basic stellar parameters of \object{HD192263} (HIP\,99711,
BD\,$-$01~3925, ADS\,13547\,A) have been discussed in detail in
\citet{San00a}. These are recalled and updated in Table\,\ref{tab1},
where we have included new spectroscopic determinations of the
atmospheric parameters \citep{San03} and of $v\,\sin{i}$
\citep{San02}.  A new $\log{R'_\mathrm{HK}}$ value is also quoted from
\citet{Hen02}.

\begin{table}
\caption[]{
\label{tab1}
Stellar parameters for \object{{\footnotesize HD}\,192263}. }
\begin{tabular}{lcc}
\hline
\noalign{\smallskip}
Parameter  & Value & Reference \\
\hline \\
$Spectral~type$  & K2V & colour-index; $M_v$  \\
$Parallax$~[mas]  & 50.27 $\pm$ 1.13 & Hipparcos \\
$Distance$~[pc]  & 19.9 & Hipparcos \\
$m_v$  & 7.79 & Hipparcos \\
$B-V$  & 0.938 & Hipparcos \\
$M_\mathrm{v}$  & 6.30 & -- \\
$Luminosity~[L_{\sun}]$  & 0.34 & \citet{Flo96} \\
$Mass~[M_{\sun}]$  & 0.75 & \citet{San03} \\
$\log{R'_\mathrm{HK}}$ & $-$4.56         & \citet{Hen02} \\
$v\,\sin{i}$~[km\,s$^{-1}$] & 1.99 & {\footnotesize CORALIE}\\
                            &      &  \citet{San02} \\
$T_\mathrm{eff}$~[K]  & 4995$\pm$50 & \citet{San03} \\
$\mathrm{[Fe/H]}$  & 0.04$\pm$0.05 & \citet{San03} \\
$\log{g}$  & 4.76$\pm$0.15 & \citet{San03} \\
\hline
\noalign{\smallskip}
\end{tabular}
\end{table}

\section{Radial velocities}
\label{rv}

\object{HD\,192263} is part of the Geneva extra-solar planet search
programme with the CORALIE spectrograph \citep[on the 1.2-Swiss
telescope, La Silla, ESO, Chile;][]{Udr00}. In this context, it was
found to present a periodic radial-velocity signal, interpreted as an
indication of the presence of a planetary mass companion orbiting this
K dwarf star \citep{San00a}. Before discussing in the next sections
the origin of the observed radial-velocity variations, let us first
simply consider the planetary explanation.

Since the planet discovery paper, we have been continuously adding
radial velocities of this star, using the CORALIE spectrograph,
gathering so a total of 182 observations. The velocities were
computed using a weighted cross-correlation mask \citep[][]{Pep02b},
which permitted to effectively reduce the rms of our measurements.

\begin{figure}[t]
\psfig{width=\hsize,file=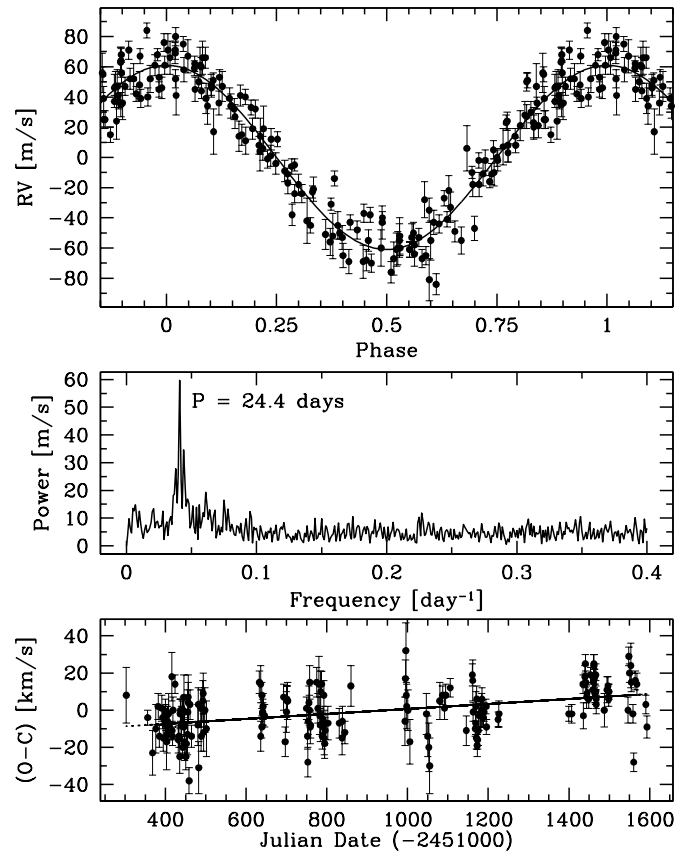}
\caption[]{{\it Upper panel}: Phase-folded diagram of the radial 
  velocities of \object{HD\,192263}. The solid curve represents the
  best Keplerian fit. {\it Middle panel}: Periodogram of the radial
  velocities, showing a very well defined peak at the observed
  $\sim$24.4-day period. {\it Lower panel}: Residuals of the 24.4-day
  orbital solution, showing the presence of a long term trend in the 
  data. The line represents a linear fit,
  and has a slope of 4.8$\pm$0.8 m\,s$^{-1}$\,yr$^{-1}$}
\label{figphase}
\end{figure}

\begin{figure*}[t]
\psfig{width=\hsize,file=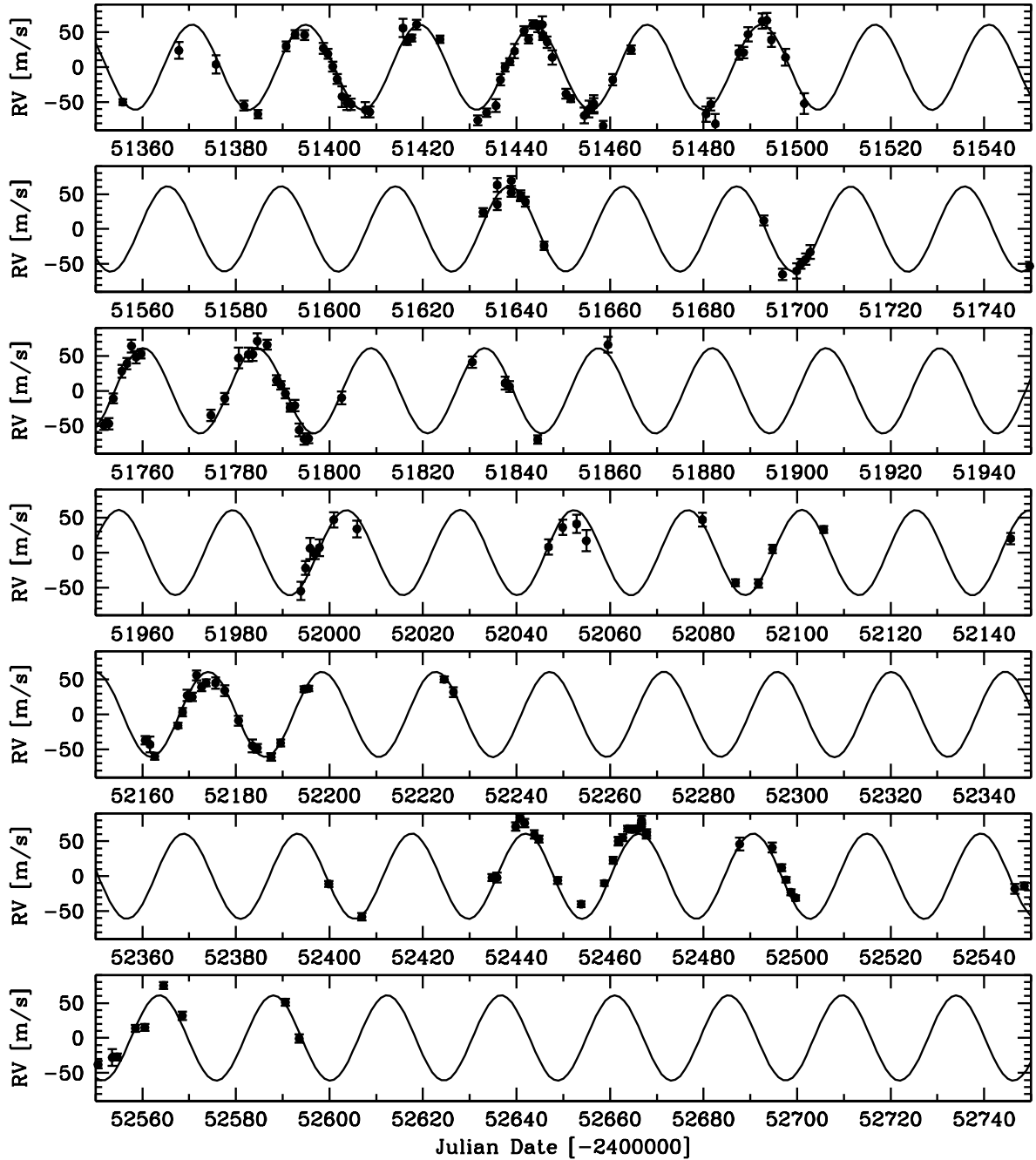}
\caption[]{Radial-velocity time series of \object{HD\,192263} for the
  complete span of our measurements. The curve represents the fitted
  orbital solution. It is interesting to see the long-term phase
  stability of the radial-velocity signal}
\label{figvr}
\end{figure*}

As before, an analysis of the radial velocities shows the presence of
a signal with a period of about 24 days and an amplitude of
61\,m\,s$^{-1}$.  In Fig\,\ref{figphase} we plot a phase-folded
diagram as well as a Fourier transform (FT) of the radial velocities.
In Fig.\,\ref{figvr} we show a time series of the observed radial
velocities for the whole period of our measurements.  The two plots
clearly show the presence of a very stable periodic radial-velocity
signal that can be interpreted as the signature of a
0.72\,M$_{\mathrm{Jup}}$ planetary companion orbiting
\object{HD\,192263} every 24.348 days, on a (quasi-)circular orbit.
The separation is about 0.15\,AU. The inferred planetary orbital
parameters and minimum mass are listed in Table\,\ref{tab2}.

Considering the stellar rotational period of 24.5\,days found by
\citet{Hen02}\footnote{A value compatible with the measured activity
  level of the star, that implyes a P$_{\mathrm{rot}}$$\sim$21\,days
  \citep{Noy84}; however, and as we will see in
  Sect.\,\ref{interactions}, it is not completely clear if the
  24.5-day photometric period is really related with the rotation
  period of the star.}, and knowing that the $v\,\sin{i}$ of
\object{HD\,192263} is 1.99\,km\,s$^{-1}$, we can estimate the orbital
inclination angle. Taking the radius of this K dwarf to be
0.8\,R$_{\odot}$, the rotational period implies a
$v_{eq}$\,$\sim$\,1.65\,km\,s$^{-1}$, very close to the observed
(minimum) value of 1.99\,km\,s$^{-1}$; both numbers are compatible
within the errors.  In other words, this star is probably seen almost
equator-on, and if the stellar rotation axis is perpendicular to the
planetary orbital plane, the measured minimum mass of the planet is
probably not far from the real mass.

\begin{table}[t]
\caption[]{
\label{tab2}
Orbital elements of the fitted orbit and main planetary properties. }
\begin{tabular}{lr@{\,$\pm$\,}ll}
\hline
\noalign{\smallskip}
$P$              & 24.348              &0.005    &[d] \\
$a_1\,\sin i$    & 0.0203              &0.0004   &[Gm]\\
$T$              & 2451979.28          &0.08     &[d] \\
$e^\dagger$      & \multicolumn{2}{c}{0.0}       & \\
$V_r$            & $-$10.686           &0.001    &[km\,s$^{-1}$]\\
$\omega^\dagger$ & \multicolumn{2}{c}{0.0}       &[deg] \\ 
$K_1$            & 61                  &1        &[m\,s$^{-1}$] \\
$f_1(m)$         & 5.64                &0.38     &[$10^{-10}\,M_{\odot}$]\\ 
$\sigma(O-C)$    & \multicolumn{2}{c}{12.5}      &[m\,s$^{-1}$]  \\    
$N$              & \multicolumn{2}{c}{182}       &   \\
$m_2\,\sin i$    & \multicolumn{2}{c}{0.72}      &[M$_\mathrm{Jup}$]\\
$a$              & \multicolumn{2}{c}{0.15}      &[AU]\\
\noalign{\smallskip}
\hline
\end{tabular}
\\ $^\dagger$ fixed; when free, $e$\,=\,0.013\,$\pm$\,0.022,
consistent with a circular orbit according to the \citet{Luc71} test.
\end{table}

As already discussed in \citet{San00a}, the residuals around the fit
are a bit higher than usual. For an average 8\,m\,s$^{-1}$ precision
for the individual measurements, the measured 12.5\,m\,s$^{-1}$
represent an excess of about 10\,m\,s$^{-1}$. The high activity level
observed for this K dwarf could explain at least part of this noise
\citep[e.g.][]{Saa98,San00b}. Using Eq.\,(1) in \citet{Saa97}, and
considering the stellar $v\,\sin{i}$ of 1.99\,km\,s$^{-1}$ and a spot
filling factor of about 1\% \citep[as found by][]{Hen02}, we estimate
that the observed radial-velocity variation should have a
semi-amplitude of about 12\,m\,s$^{-1}$. We note, however, that this
result is some sort of a maximum value, since it is computed for an
equatorial spot in a star seen equator-on.

In the lower panel of Fig.\,\ref{figphase} we present the residuals of the $\sim$24.4-day
Keplerian fit. As it can be seen from the plot, there seems to exist a long term
trend in the data, with a significant slope of about 4.8$\pm$0.8\,m\,s$^{-1}$\,yr$^{-1}$.
The source of this trend is still not clear, and might be due e.g. to the presence of
another planetary or stellar companion, or to some long-term activity-induced 
radial velocity variation connected to a possible stellar magnetic 
activity cycle \citep[e.g. ][]{Kur03}. In any case, it is in part responsible 
for the residuals of the short period fit.

Finally, we have correlated the CORALIE spectra using a Cross-Correlation mask
specially constructed for the radial-velocity determination of M4 dwarfs 
\citep[][]{Del98}. As seen for HD\,41004 \citep[][]{San02}, if the companion
to HD\,192263 was a low mass star (e.g. an M dwarf) the amplitude of the
radial-velocity signal would be dependent of the mask used. The results of our 
analysis reveal, however, that the fitted orbital parameters always remain 
unchanged. This confirms that HD\,192263 has really a very low mass companion.

\section{Planet or spurious activity signal?}
\label{spurious}

As seen in the previous section, the radial-velocity periodic signal
presented by \object{HD\,192263} has remained perfectly constant for
the last few years, showing no significant phase or amplitude
variations. Although this strongly supports the planetary explanation,
a quite similar situation was also found for \object{HD\,166435} by
\citet{Que01}. This latter case and the one presented here are,
however, quite different.  First, the two stars have different
spectral types: late-F and K dwarf, respectively. Secondly, the
rotational velocities are very different;
$v\,\sin{i}$=7.6\,km\,s$^{-1}$ for \object{HD\,166435} against
$v\,\sin{i}$=1.99\,km\,s$^{-1}$ for \object{HD\,192263}. Furthermore,
in our case the fitted ``orbit'' is perfectly compatible with
circular, contrarily to the situation for \object{HD\,166435}.
Finally, in the case of \object{HD\,166435}, there was a clear
correlation between the radial velocity and bisector measurements, not
found by \citet{San00a} for \object{HD\,192263}.

\begin{figure}[t]
\psfig{width=\hsize,file=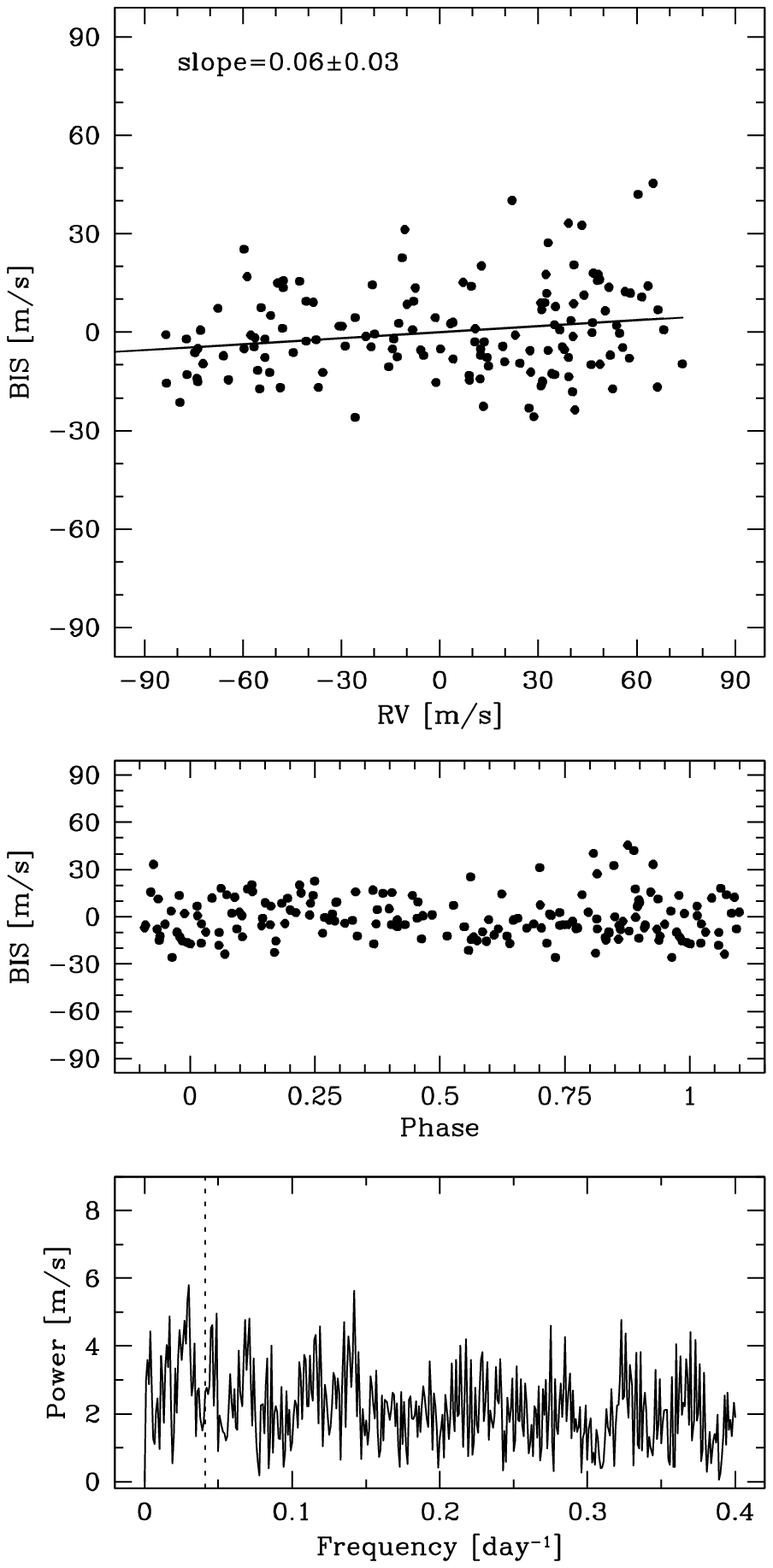}
\caption[]{{\it Upper panel}: Radial velocity vs. BIS for
  \object{HD\,192263} (as defined in \citet{Que01}). The slope and its
  uncertainty are indicated. The Spearman correlation coefficient
  between the two variables is 0.18. {\it Middle panel}: BIS values
  phase folded with the orbital period. {\it Lower panel}: FT of the
  BIS values. No significant period is found in the data. The dotted
  line is positioned at the period of 24.4-days.  Only the best
  measurements (with errors lower than 10\,m\,s$^{-1}$) are considered
  in these three plots}
\label{figbis}
\end{figure}

\subsection{Bisector analysis}
\label{bisector}

The use of the bisector analysis has been shown to be crucial in
disentangling planetary signatures from spurious radial-velocity
signals \citep[e.g.][]{Que01,San02}.  For \object{HD\,192263} this
analysis has already been presented in \citet{San00a}, revealing no
traces of bisector variations related with the radial velocity.  This
result was considered as a strong evidence for the planetary origin of
the radial-velocity signal.  However, given the doubts raised by the
recent work of \citet{Hen02}, and the much larger number of points
available now, it is worth repeating the test.

Using the procedure presented in \citet{Que01} we have computed the
Bisector Inverse Slope (BIS) for each of the measured CORALIE
cross-correlation functions (CCF's).  In Fig.\,\ref{figbis} we show
the results, plotting the derived values of BIS against the observed
radial-velocities (upper panel).  The plot shows that there is no
evident correlation between the two variables. A Spearman correlation
coefficient of 0.18 is obtained. Thus, BIS does not significantly
correlate with the radial velocities.

It is important to further note that the observed slope has the
opposite sign than in the case of \object{HD\,166435}, for which an
anti-correlation was found.  \object{HD\,166435} is the only clear
published example of activity-induced planetary-like signature. Other
cases from the {\small CORALIE} planet-search programme are under
study \citep[see e.g. ][]{San00c}.

In the two lower panels of Fig.\,\ref{figbis} we plot a phase-folded
diagram of the BIS, constructed using the same period as observed in
the radial velocity, as well as the FT of the data. No periodic
variation seems present. In particular, we see no sign of variations
of BIS with the period of $\sim$24.4-days, as seen in the
radial-velocity data. These results strongly support the planetary
explanation proposed by \citet{San00a} and \citet{Vog00} as the source
for the radial-velocity signal.

\subsubsection{Efficiency of the bisector diagnostic for low rotators?}

Using a simple model, we have checked the sensitivity of the bisector
analysis to discriminate activity-related radial-velocity variations
from real planetary signatures.  Our model consists in a stellar disk
divided in a grid of 200$\times$200 cells.  Considering that the
stellar-template CCF (i.e. a spectral line) in the center of the
stellar disk is well approximated by a Gaussian function with given
depth and width\footnote{values of 0.25 and 4.30\,km\,s$^{-1}$ are
  respectively chosen as typical values for {\small CORALIE} spectra.},
we have computed the CCF for each cell. In this process we took into
account the cell positions on the disk, to account for the
limb-darkening effect \citep[a factor of 0.6 was taken -- ][]{Gra92}, and 
the projected radial velocity. The different CCF's were then
added. This procedure was repeated for several ``stars'' with
different $v\sin{i}$ values (from 0 to 10.0\,km\,s$^{-1}$).

From the resulting stellar-disk CCF's, we have subtracted a CCF
corresponding to the light that is masked by an equatorial stellar
spot positioned at an angle of 65 degrees from the center of the disk
(the star is seen equator-on). This value was taken to coincide with
the approximate maximum perturbation in radial velocity that a spot
might produce \citep[see e.g. Fig.\,1 in][]{Saa97}.  Spots with
different filling factors were considered as well.

\begin{figure}[t]
\psfig{width=\hsize,file=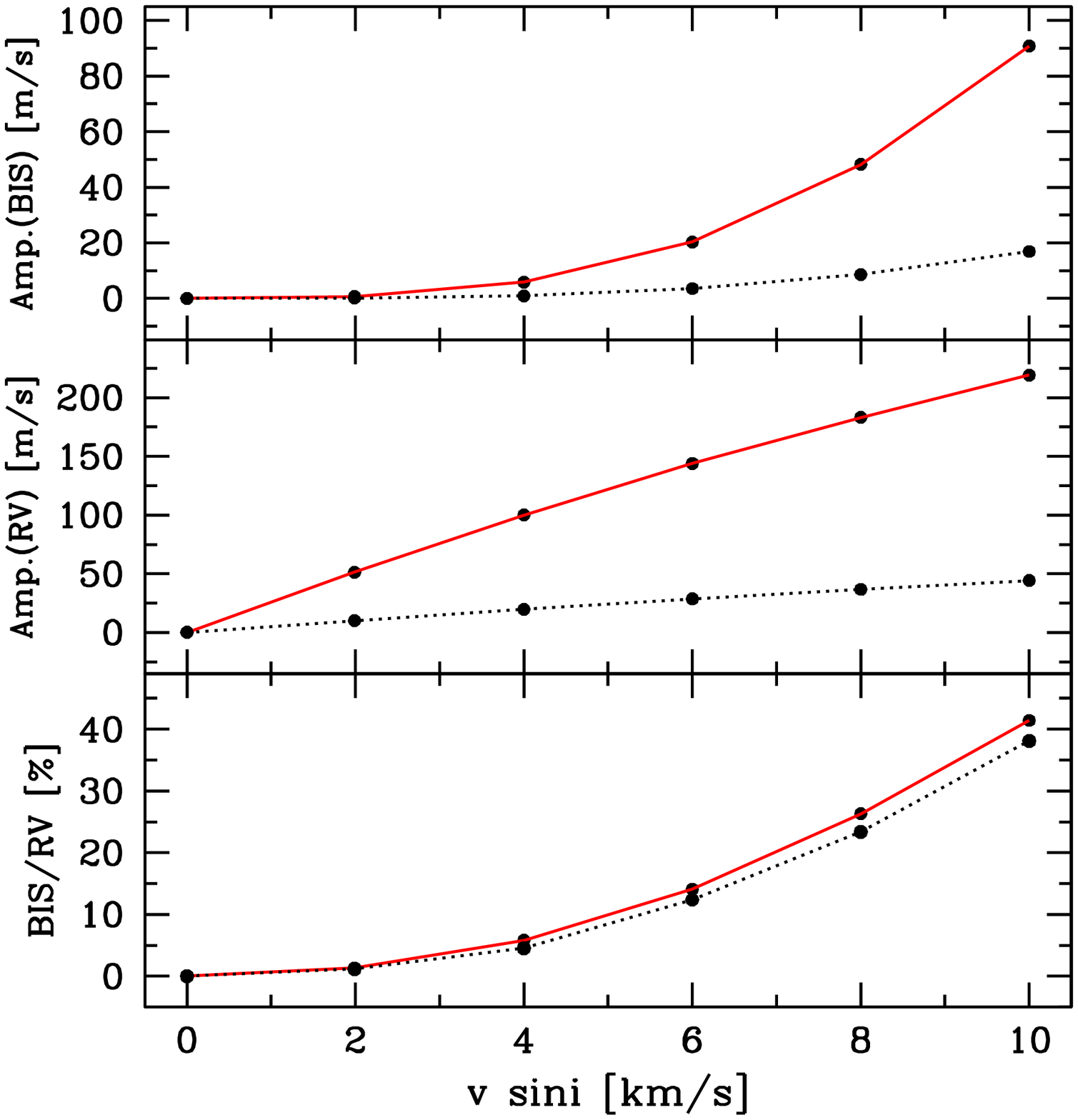}
\caption[]{{\it Upper panels}: Modeled amplitudes of bisector inverse
  slope (BIS) and radial velocity, induced by spots with filling
  factors of 5\% and 1\% (solid and dotted line, respectively), 
  plotted as a function of the star's projected rotational
  velocity. {\it Lower panel}: Amplitude ratios (in percent) as a
  function of $v\,\sin{i}$. These results were obtained with a very
  simple model, and should be considered only as qualitative.
  See text for more details }
\label{figmodel}
\end{figure}

The results of this simple model are presented in Fig.\,\ref{figmodel}
where we plot the radial velocities, bisector 
``span'' \citep[in this case the Bisector Invese Slope, BIS, as defined in][]{Que01}, and the ratio of the BIS-to-radial-velocity amplitudes, 
as a function of $v\,\sin{i}$. We
observe that, for low $v\sin{i}$ values, the influence of a spot on
the measured velocity is much larger than the effect observed on the
CCF bisector. In the case of {\object HD\,192263}, with
$v\sin{i}$=1.99\,km\,s$^{-1}$, this ratio is around 1-2\%. The
radial-velocity (semi-)amplitude induced in this low $v\sin{i}$ model
for a spot with a filling factor of 5\% (larger than the one observed
in photometry for \object{HD\,192263}) is around 50\,m\,s$^{-1}$ i.e.
similar to the actually observed value. For the same model we find,
however, a BIS with an amplitude of only $\sim$1\,m\,s$^{-1}$.  Both
values are in good agreement with the results obtained from Eqs.\,(1)
and (2) in \citet{Saa97}, attesting the reality of this effect. 
However, some differences exist since the bisector span definition used by
these authors is not the same as the one used here, and probably also
due to the simplification of our model.

As it was also shown by \citet{Saa97}, for higher $v\sin{i}$ models
this ratio is higher. The effect is of the order of $\sim$50\% for a
$v\sin{i}$=10.0\,km\,s$^{-1}$, and thus clearly detectable. This is
mostly due to the higher sensitivity on $v\sin{i}$ of the effect the
spot has onto the bisector when compared with the effect induced on
the velocity itself \citep{Saa97}.  
As shown by \citet[][]{San00c}, this qualitative effect is indeed observed, 
as there seems to exist a clear correlation between the observed 
BIS-to-radial-velocity amplitude ratios and the stellar projected rotational 
velocity.

The model described above is, however, far from
being perfect. For example, we would obtain a value for the BIS-to-radial-velocity
ratio of only $\sim$0.2 in the case of \object{HD\,166435} (a late-F dwarf with
$v\sin{i}$=7.60\,km\,s$^{-1}$), while the observed value is close to 1 \citep{Que01}.  
The difference is probably due to effects not taken
into account, connected e.g. to the geometry of the system or even to
the presence of inhomogeneous convection effects -- see discussion in
\citet[][]{Saa97}. The results obtained here thus probably represent lower
limits for the bisector variations. Finally, changes in the spot
filling factor do not seem to strongly influence the
BIS-to-radial-velocity amplitude ratio.

In other words, it is possible that the bisector test is less
sensitive for slow rotators, although a small effect should still be
visible\footnote{Although not the case with the CORALIE spectra (R=50\,000), 
we should add that the spectral resolution might also impose limits to the 
validity of the bisector test. In fact, given that the broadening factors
essentially sum up in quadrature, it is very difficult to put in evidence
intrinsic line asymmetries when the instrumental profile is
significantly broader than the intrinsic line profile.}. But we further 
caution that this very simple model 
has its own limitations, and the effects discussed above should thus be 
seen as a qualitative but not quantitative result. The derived BIS-to-radial-velocity amplitude ratios should not be taken as 
established values. Finally, these facts do not exclude the planetary explanation for the case of \object{HD\,192263},
but rather show that for this particular star the bisector test may not be
as efficient as we could eventually imagine. Another diagnostic is
needed.

\subsection{Photometry}
\label{fotommetry}

The results presented in \citet{Hen02}, calling into question the
planetary nature of the observed radial-velocity variations, are
mostly based on the discovery that \object{HD\,192263} has a periodic
photometric signal with a period similar to the one observed in
radial-velocity. However, \citet{Hen02} could not directly compare the
photometry and the radial-velocities obtained at the same moment in
time. This comparison should be done to completely establish a
relation between the origin of these two quantities (photometry and
radial velocities).

In order to address this problem, while monitoring the radial
velocities of \object{HD\,192263} with {\small CORALIE}, we have
started a simultaneous photometric campaign on this star.  From May 30
to November 5, 2002, \object{HD\,192263} has been measured 187 times
in the \textsc{Geneva} photometric system \citep{golay} with the
photoelectric photometer P7 \citep{burnet}, completely refurbished in
2001 and mounted on the 120-cm Belgian \textsc{Mercator} telescope in
La~Palma (IAC, Canary Islands, Spain). The global photometric
reduction procedure is described in Rufener \citep{rufener1,rufener2}.
However, in this particular case, two additional comparison stars,
\object{HD\,194953} (G8~III) and \object{HD\,196712} (B7~III) have
been systematically measured together with \object{HD\,192263}, in
order to improve the final data set.  The photometric data in the
\textsc{Geneva} system are collected in the General Catalogue \citep{rufener3}
and its up-to-date database \citep{burki4}.  An analysis of the
standard stars shows that the final precision of the data is between
0.002 and 0.004\,mag\footnote{Besides this photometric campaign, C.
  Nitschelm (private communication) has furnished us a series of
  photometric measurements of \object{HD\,192263} obtained during the
  last three years at the Danish 0.5-m telescope (La Silla, ESO,
  Chile). The data show no special photometric variations for this
  star over the whole period.  However, given the obtained precision
  of only about 0.01\,mag, these observations cannot be used to
  strongly constrain our results, and in particular to check for the
  presence of short period photometric variations.}.  
  
Finally, we have also used the photometric measurements published 
in \citet{Hen02}, as we have radial-velocity measurements that 
coincide (in time) with these data.
 
A look at the \textsc{Geneva} data reveals that, indeed, there are clear
photometric variation as observed by \citet{Hen02}. However, these
variations do not seem to be stable. We shall discuss the different
aspects of the question in the following subsections.

\subsection{Correlating the various parameters}
\label{correlating}

\begin{figure}[t]
\psfig{width=\hsize,file=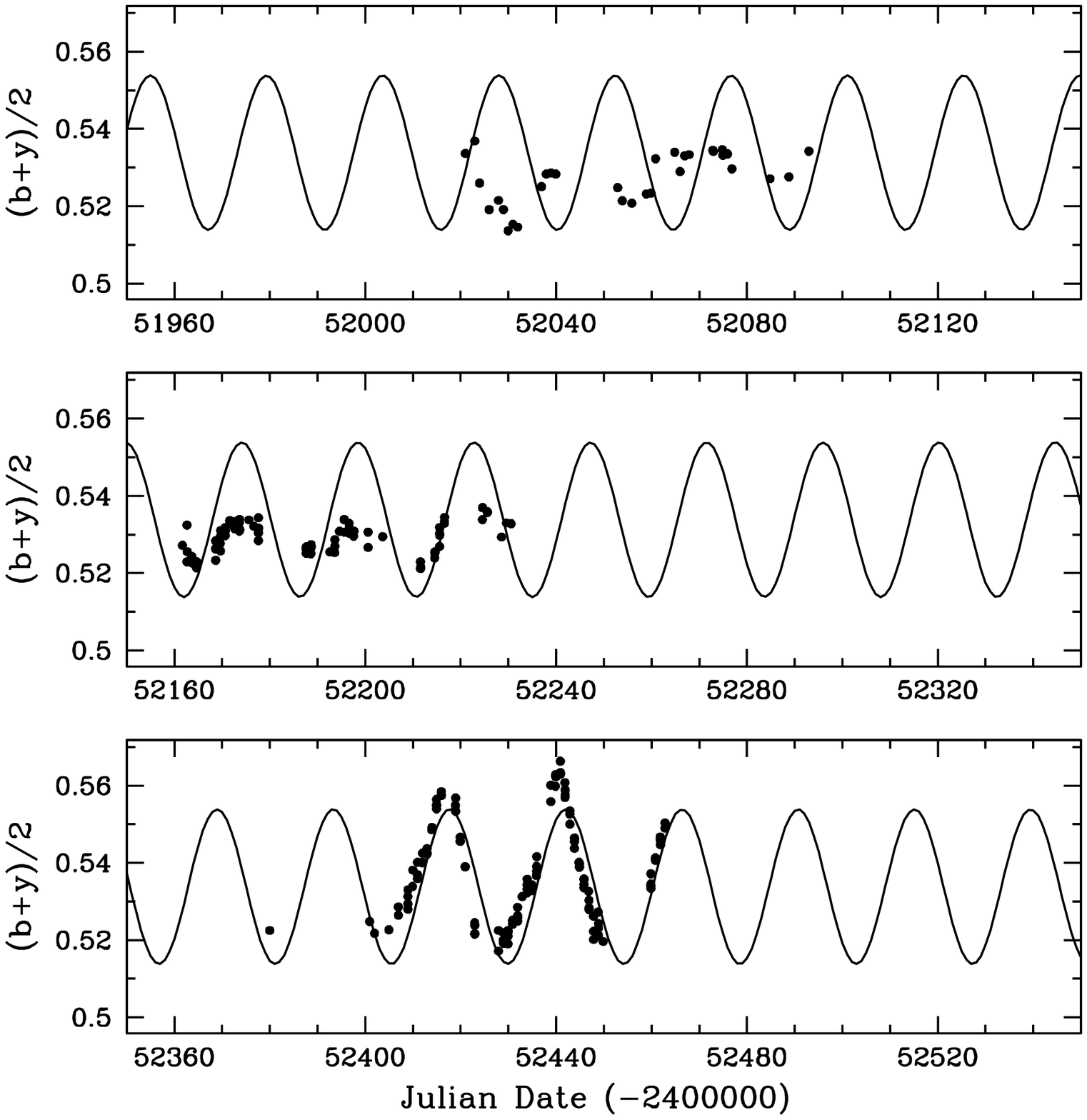}
\caption[]{Photometric measurements of \citet{Hen02} plotted as 
  a function of time. The three panels correspond to panels 4, 5 and 6
  of Fig\,\ref{figvr}.  Superimposed with the photometry is a
  sinusoidal curve with the same period and phase as the
  radial-velocity Keplerian fit}
\label{fighenry}
\end{figure}

In Fig.\,\ref{fighenry} we can see a plot of the photometric
measurements listed by \citet{Hen02}. These data were used by the
authors to show that \object{HD\,192263} presents a periodic
photometric variation with the same period as the one observed in
radial velocity. In the diagram, we have drawn a sinusoidal curve with
the same phase and period as the one observed in radial-velocity (same
as seen in Fig.\,\ref{figvr}). The amplitude of this curve was
visually set to reproduce the amplitude of the last group of points.

A look at the figure reveals two interesting features. First, for the
last measurements, there is a clear superposition of the
``radial-velocity'' curve and the photometry; the phase and period
seems to be about the same. However, for the earlier cycles (two upper
panels) this fact is not present. Although the precision of the
measurements is the same (as can be seen from the very small
dispersion), the amplitude is much smaller.  This is, of course, in
complete disagreement with the stability of both phase and amplitude
observed in the radial velocities. For example, in the period between
JD=2\,452\,160 and 2\,452\,200, for which both RV and photometry
exist, the latter looks almost stable while the radial velocities vary
with the usual ($\sim$60\,m\,s$^{-1}$) amplitude, period, and phase (see 
Fig.\,\ref{figvr}). Furthermore, in the upper panel of Fig.\,\ref{fighenry}, 
it seems that for the region around JD=2\,452\,040 the phase of the 
photometric variations has changed.

These considerations already strongly suggest that, during the time of
the photometric measurements, either there has been a phase shift in
the position of the spot group (eventually disappearing an
re-appearing at another location) or else that the stellar rotational
period (responsible for the photometric variation) is not exactly the
same as the one found in radial velocity.  In any case, these
observations seem to be contradictory with the idea that the observed
radial-velocity variation is being induced (only) by the presence of
the spots responsible for the photometric variations.

\begin{figure}[t]
\psfig{width=\hsize,file=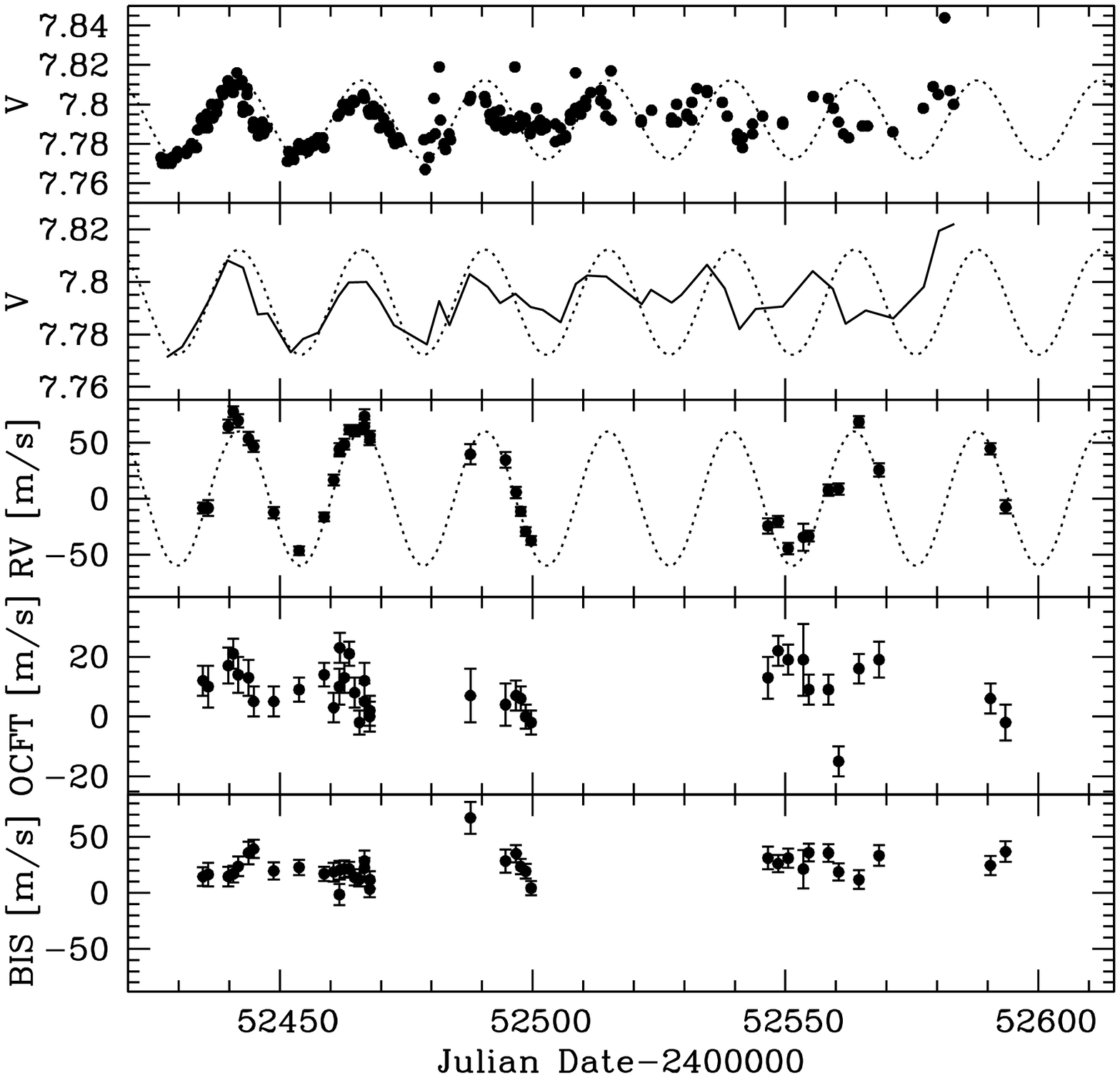}
\caption[]{Plots of the \textsc{Geneva}-photometry points (upper panel) and
  3-day binned photometric data (second panel, solid line), {\small CORALIE} 
  radial velocities, residuals to the Keplerian fit (OCFT), and BIS values,
  as a function of time for the period of simultaneous observations.
  The sinusoidal curve on top of the velocity points (third panel)
  illustrates the global Keplerian solution obtained for our data (see
  Sect.\,\ref{rv}).  For the two panels with the photometric data, 
  the ``fitted'' dotted line has the same period and phase as for the radial 
  velocities}
\label{figcomp}
\end{figure}

To further investigate this fact, we plot in Fig.\,\ref{figcomp} the
simultaneous temporal sequencies of \textsc{Geneva}-photometry observations,
{\small CORALIE} radial velocities, residuals to the Keplerian
24.4-day fit, and BIS values. In the three upper panels, we have drawn a
sinusoidal curve corresponding to the best Keplerian solution derived
from the radial-velocity data (see Sect.\,\ref{rv}).  As previously,
for the photometry, the amplitude of this curve was adjusted in order
to better fit the data.

The comparison of these plots shows that while the radial velocities
follow a very stable (in period, amplitude, and phase) periodic
variation, the photometry presents a strange behaviour. At first, we
see a clear variation with the same period and phase as the radial
velocities.  However, from a given moment on, the photometric
variations become quite random, and no clear periodicity exists any
more. There is even the impression that the relative phase of the 
photometric and radial-velocity variations is changing slowly, 
something that could imply e.g. that the rotational period of the
star and the radial-velocity periodic signal do not have the very same length.
During the whole period, both BIS and residuals are reasonably
constant in time. A look at the residuals (mostly for the first group
of points) shows, however, what seems to be a small amplitude
(about 20\,m\,s$^{-1}$) coherent radial-velocity signal
left\footnote{The quality of the data does not permit to precisely
  access the phase of this periodic signal.}. The same marginal trend is
seen for the BIS during this particular time interval for which the
photometry shows a clear periodic variation. These trends might be
related to the radial-velocity variation induced by the spot group.
But the fact that in the global residuals (see Sect.\,\ref{rv}) no
similar period appears in the Fourier Transform, suggests that this
variation is sporadic, as expected since the photometric variations
are not stable\footnote{We have tried to verify if there was any
  relation between the radial-velocity residuals and the BIS. Nothing
  is seen, maybe because of the large errors in the individual
  measurements, when compared to the magnitude of the effect.}.

This simultaneous analysis brings many doubts onto the conclusions of
\citet{Hen02}. If the presence of spots (and other stellar surface
features) were the source for the radial-velocity variations, we
should definitely see a correlation between the photometry and the
radial-velocity data for the period of our simultaneous measurements.
Nothing is seen.  Except if we imagine that there are other stellar
features not observable in photometry but on the other hand able to
induce radial-velocity variations, it is very difficult to accept that
the observed stable radial-velocity signal is being caused by
activity-related phenomena\footnote{In this context, note also that
  for K dwarfs the convective velocities are not very high (smaller
  than for F dwarfs), reducing the probability of radial-velocity
  variations induced by convective inhomogeneities
  \citep{Saa97,Saa98,San00b}.}.

Another interesting detail also deserves some attention. If a spot is
responsible for a significant radial-velocity signal, there should be
a phase shift between the photometry and the radial-velocity signal.
This shift, observed by \citet{Que01} for \object{HD\,166435}, may
be justified by simple considerations. When a (single) spot is at the
center of the disk, the photometric ``variation'' should be the
highest. At the same time, the spot will cause no radial-velocity
shift.  On the other hand, the maximum effect in radial velocity
should happen when the spot is located in an intermediate position
between the center of the disk and the limb \citep[e.g.][]{Saa97}.
According to this simple view, the fact that in the first group of
points in Fig.\,\ref{figcomp} we see a clear phase-alignment between
the radial-velocity and the photometric measurements, suggests that
these two quantities cannot be directly related.

\subsection{Star-planet interactions?}
\label{interactions}

As seen in the previous sections, the idea that the observed
radial-velocity periodic variations are being induced by the presence
of photospheric features (e.g. spots) is not satisfying as a whole. In
other words, the planetary model, as first discussed in \citet{San00a}
and \citet{Vog00} should again be considered as the best explanation.
Still, the fact that the rotational period of the star seems,
according to the photometry, to be similar
to the planetary orbital period is intriguing.

One simple way of explaining this is to say that this is pure
coincidence, i.e. that the rotational period of the star is by chance
similar (but not necessarily equal) to the planetary orbital period.
This is not so unrealistic. Known exoplanets have periods from a few
days to several years. Most of the stellar low rotators (as e.g. the
Sun) have rotational periods of the order of 20 days. They, moreover,
form the sample bulk of the programmes searching for planets with the
radial-velocity technique. The probability that a 24-d period planet
fall in this subsample of low rotating star is large. Of course, the
closer the photometric and orbital periods, the lower this
probability.

Another possibility can also be explored. In the last couple of years
a few studies have been published regarding the interaction between
the exoplanets and their host stars. As discussed by \citet{Cun00} and
\citet{Saa01}, this interaction might be the result of tidal or/and
magnetic effects. These can thus be responsible for observable
features such as chromospheric/coronal heating, or chromospheric
activity phenomena, possibly inducing changes in the measured radial
velocities.  A star suffering strong magnetic interactions with the
planet could e.g. have a {\sl hot spot} rotating with the same period
as the planet. On the other hand, tidal interactions could induce
variations with half the orbital period. 
In this context, \citet{Rub00} have proposed that the observation 
of ``superflares'' might be related to the presence of close-in giant planets. 
Recently, \citet{Shk02,Shk03} have found evidences that a few stars known to
harbour close-in planetary companions present important activity
induced effects, observable as variations in the \ion{Ca}{ii} H \& K
line reversals.  Based on this discovery, we suggest that the
photometric variability observed for \object{HD\,192263} might very
well be the result of such kind of star-planet interactions.

As discussed in Sect.\,\ref{correlating}, there is no offset in phase
between the radial-velocity and the photometric variations (over the
period of time during which the two persist). This shift means that the
spot is at the ``center'' of the disk (and unable to
induce radial-velocity variations) when the planet is producing the 
maximum radial-velocity variation (i.e. when it is located at phase 0.0). 
In other words, the spot is not located at the sub-planetary point, but
rather at an angle of about 90 degrees.
This fact could be seen a signature that the spot (or spot group) is due 
to tidal friction effects, and that the delay is due to a process similar to 
the ones producing the phase shift between the position of the moon and the 
maximum tidal height on earth\footnote{It should be noted, however, that the
  timescales of the tidal motions in the low-density stellar
  atmosphere are short \citep{Cun00}.} (or to some other physical process causing 
a delay in the reaction of the stellar photosphere/chromosphere to the perturbations).
Such an offset was also possibly found by \citet{Shk02} for $\upsilon$\,And.
If true, this could also mean that the system is not synchronized; as we have seen
in the previous sections, this would not be a surprise, since there might be a continuous
phase variation between the radial-velocity and photometric signals. 
It is further interesting to note that a period half of the orbital one
is seen in the Mount Wilson S index \citep{Hen02}, a possible value if
the effects observed are of tidal origin \citep{Cun00}. 

However, the current data do not give us the possibility to exclude any of the
possibilities (tidal, magnetic, or others). In the case of a magnetic
origin for the interaction, this would give us the possibility to
access the magnetic field of the planet, which would be, of course, an
interesting result.
 
\section{Conclusions}
\label{conclusions}

We have presented the case of \object{HD\,192263}, a star that shows
very stable periodic radial-velocity variations. While these were
first considered as the signature of a planetary companion, this
interpretation was recently called back into question by the detection
of periodic photometric variations with a period very close to the
orbital period \citep{Hen02}.  In order to understand the true source
of the observed radial-velocity signal, we have gathered precise
photometry, radial-velocity, and bisector measurements for this star,
part of them taken simultaneously.  Our results are the followings:
\medskip

-- The radial-velocity variations show a striking long-term stability
in period, phase, and amplitude. A very different scenario is found
for the photometry that alternates moments of ``stability'' with
periods of variability.  Cross-correlation BIS (Bisector Inverse
Slope) measurements do not correlate either with radial velocities.
These observations very strongly supports the idea that photometry and
radial-velocity variations do not share the same origin.  Except if
some unknown photospheric phenomenon is being responsible for the
observed radial-velocity variation without influencing the stellar
photometric behaviour in the same manner, the $\sim$24.4-day period in
the radial-velocity data remains best interpreted by the presence of a
low-mass planet around HD\,192263.

-- The similarity of the measured radial-velocity and photometric
periods can be interpreted in several ways. On the one hand, it can be
a simple coincidence. On the other hand, it can be the result of
interactions (magnetic or tidal) between the planet and the star, and
able to induce activity-related phenomena.

-- The radial-velocity data also show the possible signature of 
a long-term trend.
The nature of this trend is not known. Possible explanations might
involve the presence of another planetary (or stellar) companion, or even
long term activity-induced radial velocity variations. A clear followup 
of this result will be possible with instruments like HARPS \citep{Pep02}.  
\medskip

This paper shows that the use of photometry as a tool to confirm (or
not) the presence of the planetary mass companions to solar-type
stars, detected by radial-velocity technique, should be taken
cautiously. The results should always be analyzed carefully, and
whenever possible, the various data (radial-velocities, photometry,
bisector) should be obtained simultaneously in time.

\begin{acknowledgements}
  Nuno Santos dedicates this work to the memory of Gustavo Camejo Rodrigues, who will 
  always be dearly remembered. We would like to thank Tim Brown for fruitful discussions, 
  C.\,Nitschelm, A.\,Vidal-Madjar, and A.\,Lecavelier for having provided us
  with photometric measurements of HD\,192263, and the numerous observers at 
  the \textsc{Mercator} telescope, from Leuven and Geneva, for the successful 
  photometric monitoring. We wish to thank the Swiss National Science Foundation 
  (Swiss NSF) for the continuous support to this project. Support 
  from Funda\c{c}\~ao para a Ci\^encia e Tecnologia, Portugal, to N.C.S., in the form of a
  scholarship, is gratefully acknowledged.
\end{acknowledgements}

\end{document}